\documentclass[a4paper,12pt]{article}
\usepackage{comment}
\usepackage{cite}
\usepackage{amsmath}
\usepackage{amssymb}
\usepackage{amsfonts}
\usepackage[T1]{fontenc}
\usepackage[utf8]{inputenc}
\usepackage{graphicx}
\usepackage{fancyhdr}
\usepackage{float}
\usepackage{xcolor}
\usepackage{authblk}
\usepackage{mathrsfs}
\usepackage{empheq}
\usepackage[hyphens]{url}
\usepackage{hyperref} 
\usepackage[]{breakurl}

\usepackage{geometry}
\newcommand{\threej}[6]{\left(\begin{array}{ccc}#1 & #2 & #3 \\ #4 & #5 & #6 \end{array}\right)}

\begin{document}

\title{Sister Celine's polynomials in the quantum theory of angular momentum}

\author[$\dagger$]{Jean-Christophe {\sc Pain}$^{1,2,}$\footnote{jean-christophe.pain@cea.fr}\\
\small
$^1$CEA, DAM, DIF, F-91297 Arpajon, France\\
$^2$Universit\'e Paris-Saclay, CEA, Laboratoire Mati\`ere en Conditions Extr\^emes,\\ 
F-91680 Bruy\`eres-le-Ch\^atel, France
}

\maketitle

\begin{abstract}
The polynomials introduced by Sister Celine cover different usual orthogonal polynomials as special cases. Among them, the Jacobi and discrete Hahn polynomials are of particular interest for the quantum theory of angular momentum. In this note, we show that characters of irreducible representations of the rotation group as well as Wigner rotation ``$d$'' matrices, can be expressed as Sister Celine's polynomials. Since many relations were proposed for the latter polynomials, such connections could lead to new identities for quantities important in quantum mechanics and atomic physics. 
\end{abstract}

\section{Introduction}

Sister Celine introduced the polynomial \cite{FASENMYER1945,FASENMYER1947} (see also Refs. \cite{RAINVILLE1960,ZEILBERGER1982}):
\begin{equation}\label{orig}
    f_n\left[\begin{array}{c}
    a_1, \cdots, a_p\\
    b_1, \cdots, b_q
    \end{array};x\right]=_{p+2}F_{q+2}\left[
    \begin{array}{c}
    -n, n+1, a_1, \cdots, a_p\\
    1, \frac{1}{2}, b_1, \cdots, b_q
    \end{array};x
    \right]
\end{equation}
defined ($|t|<1$) by the generating function
\begin{equation}
    \sum_{n=0}^{\infty}f_n\left[\begin{array}{c}
    a_1, \cdots, a_p\\
    b_1, \cdots, b_q
    \end{array};x\right]t^n=\frac{1}{(1-t)}~_{p+2}F_{q+2}\left[
    \begin{array}{c}
    -n, n+1, a_1, \cdots, a_p\\
    1, \frac{1}{2}, b_1, \cdots, b_q
    \end{array};-\frac{4xt}{(1-t)^2}
    \right],
\end{equation}
where $_pF_q$ denotes the generalized hypergeometric function.  The Sister Celine polynomials can be generalized in the following manner \cite{JAIN1967}:
\begin{equation}\label{jain1967}
    \mathscr{J}_n^{(c,k)}(a_1,\cdots,a_p; b_1,\cdots, b_q;x)=\frac{(c)_n}{n!}~_{p+k}F_{q+k}\left[
    \begin{array}{c}
    -n,\Delta(k-1,c+n), a_1, \cdots, a_p\\
    \Delta(k,c), b_1, \cdots, b_q
    \end{array};(k-1)^{k-1}x
    \right],
\end{equation}
where $\Delta(k,c)$ is defined as the set of parameters $c/k, (c+1)/k, \cdots, (c+k-1)/k$. The original Sister Celine polynomials of Eq. (\ref{orig}) are recovered for $c=1$ and $k=2$. One has also
\begin{equation}
    \mathscr{J}_n^{(1+\alpha+\beta,2)}\left(\frac{\alpha+\beta+1}{2},\frac{\alpha+\beta}{2}+1;1+\alpha;x\right)=\frac{(1+\alpha+\beta)_n}{(1+\alpha)_n}\,P_n^{(\alpha,\beta)}(1-2x),
\end{equation}
as well as 
\begin{equation}
    \mathscr{J}_n^{(1+\alpha+\beta,2)}\left(\frac{\alpha+\beta+1}{2},\frac{\alpha+\beta}{2}+1,\xi;1+\alpha,p;v\right)=\frac{(1+\alpha+\beta)_n}{(1+\alpha)_n}\,H_n^{(\alpha,\beta)}(\xi,p,v),
\end{equation}
where $H_n^{(\alpha,\beta)}(\xi,p,v)$ is a generalized Rice polynomial \cite{KHANDEKAR1964}, which can be related to the Jacobi polynomial by
\begin{equation}
    \sqrt{\gamma}(1+t)^{-\gamma}\,P_n^{(\alpha,\beta)}\left(1-\frac{2x}{1+t}\right)=\frac{1}{2i\pi}\int_{\sigma-i\infty}^{\sigma+i\infty}\sqrt{\lambda}\sqrt{\gamma-\lambda}\,H_n^{(\alpha,\beta)}(\lambda,\gamma,x)\,t^{-\lambda}d\lambda.
\end{equation}
Jain also established integral relations, which were extended by Khan \cite{KHAN1989} using the Mellin inversion formula. In 1970, Shah defined \cite{SHAH1970}:
\begin{equation}
    F_n(x)=x^{(m-1)n}\,_{p+m}F_q\left[
    \begin{array}{c}
    \Delta(m,-n), a_1, \cdots, a_p\\
    b_1, \cdots, b_q
    \end{array};\lambda x^{\mu}
    \right]
\end{equation}
with $\Delta(m,-n)=-n/m,(-n+1)/m,\cdots,(-n+m+1)/m$. Setting $m=\lambda=\mu=1$, one gets
\begin{equation}
    F_n(x)=~_{p+1}F_q\left[
    \begin{array}{c}
    -n, n+\alpha+\beta+1, a_2, \cdots, a_p\\
    1+\alpha, \frac{1}{2}, b_3, \cdots, b_q
    \end{array};x
    \right]
\end{equation}
and for $p=q=3$, $a_2=1/2$, $a_3=\xi$, $b_3=p$, 
\begin{equation}
    F_n(x)=~_3F_2\left[
    \begin{array}{c}
    -n, n+\alpha+\beta+1, \xi\\
    1+\alpha, p
    \end{array};x
    \right]=\frac{n!}{(1+\alpha)_n}\,H_n^{(\alpha,\beta)}(\xi,p,x)
\end{equation}
and finally for $p=q=3$, $a_2=1/2$, $a_3=b_3=\xi=p$,
\begin{equation}
    F_n(x)=~_3F_2\left[
    \begin{array}{c}
    -n, n+\alpha+\beta+1\\
    1+\alpha
    \end{array};x
    \right]=\frac{n!}{(1+\alpha)_n}\,P_n^{(\alpha,\beta)}(1-2x).
\end{equation}
Such generalized polynomials were also denoted by Khan in the slightly different form \cite{KHAN1999}:
\begin{equation}\label{khan1999a}
    f_n(k,\lambda,\mu;a_1,\cdots,a_p; b_1,\cdots, b_q;x)=~_{p+k+1}F_{q+k+1}\left[
    \begin{array}{c}
    \Delta(k,-n), n+\lambda, a_1, \cdots, a_p\\
    \Delta(k+1,\mu), b_1, \cdots, b_q
    \end{array};x
    \right],
\end{equation}
which enables one to recover the Jacobi polynomials through
\begin{equation}
    f_n(1,1+\alpha+\beta,1+2\alpha;\alpha+\frac{1}{2};-;\frac{1-x}{2})=~_{2}F_{1}\left[
    \begin{array}{c}
    -n, n+\alpha+\beta+1\\
    1+\alpha
    \end{array};\frac{1-x}{2},
    \right],
\end{equation}
i.e.,
\begin{equation}
    f_n(1,1+\alpha+\beta,1+2\alpha;\alpha+\frac{1}{2};-;\frac{1-x}{2})=\frac{n!}{(1+\alpha)_n}\,P_n^{(\alpha,\beta)}(x).
\end{equation}
Note that we have also
\begin{align}
    f_n(1,1+\alpha+\beta,1+2\alpha,\xi;\alpha+\frac{1}{2};p;v)&=~_{3}F_{2}\left[
    \begin{array}{c}
    -n, n+\alpha+\beta+1\\
    1+\alpha, p
    \end{array};v
    \right]\nonumber\\
    =&\frac{n!}{(1+\alpha)_n}\,H_n^{(\alpha,\beta)}(\xi,p,v).
\end{align}
More recently, Ahmad \emph{et al.} \cite{AHMAD2016} and \"Ozmen \cite{OZMEN2019} considered:
\begin{equation}
    \mathscr{A}_n^{(\alpha,\beta)}\left[\begin{array}{c}
    a_1, \cdots, a_p\\
    b_1, \cdots, b_q
    \end{array};x\right]=\frac{(1+\alpha+\beta)_n}{n!}~_{p+2}F_{q+2}\left[
    \begin{array}{c}
    -n, n+\alpha+\beta+1, a_1, \cdots, a_p\\
    1+\alpha, \frac{1}{2}, b_1, \cdots, b_q
    \end{array};x
    \right],
\end{equation}
giving 
\begin{equation}
    \mathscr{A}_n^{(\alpha,\beta)}(x)=\frac{(1+\alpha+\beta)_n}{(1+\alpha)_n\sqrt{\pi}}\int_0^1t^{-1}(1-t)^{1/2-1}\,P_n^{(\alpha,\beta)}(1-2xt)dt.
\end{equation}

In the following, we keep the notation of Eq. (\ref{khan1999a}) \cite{KHAN1999}.

\section{Rotation matrices and characters of irreducible representations of the rotation group}

The main idea of the present part is to take advantage of the relation
\begin{equation}
    P_s^{(\mu,\nu)}(\cos\theta)=\frac{(1+\mu)_s}{s!}\,_2F_1\left[\begin{array}{c}
    -s, s+\mu+\nu+1\\
    1+\mu
    \end{array};\frac{1-\cos\theta}{2}
    \right]
\end{equation}
yielding, with a Sister Celine polynomial
\begin{equation}
    P_s^{(\mu,\nu)}(\cos\theta)=\frac{(1+\mu)_s}{s!}\,f_s\left(1,\mu+\nu+1,1+2\mu,\mu+\frac{1}{2};-;\frac{1-\cos\theta}{2}\right).
\end{equation}

\subsection{Characters of irreducible representations}

The character of the irreducible representation of rank $j$ of the rotation group can be put in the form \cite{VARSHALOVICH1988}:
\begin{equation}
\chi^j(\omega)=\frac{(4j-2)!!}{2(4j+1)!!}\,P_{2j}^{(1/2,1/2)}\left[\cos\left(\frac{\omega}{2}\right)\right]
\end{equation}
yielding
\begin{align}\label{chij}
    \chi^j(\omega)=&\frac{(4j-2)!!}{2(4j+1)!!}\frac{\left(\frac{3}{2}\right)_{2j}}{(2j)!}\,_2F_1\left[\begin{array}{c}
    -2j,2j+2\\
    3/2\\
    \end{array};\sin^2\left(\frac{\omega}{4}\right)\right].
\end{align}
or, in terms of Sister Celine's polynomials
\begin{align}\label{chijss}
    \chi^j(\omega)=&\frac{(4j-2)!!}{2(4j+1)!!}\frac{\left(\frac{3}{2}\right)_{2j}}{(2j)!}\,f_{2j}\left(1,2,2;1;-;\sin^2\left(\frac{\omega}{4}\right)\right).
\end{align}
Note that we have also
\begin{equation}
    P_n^{(1/2,1/2)}(x)=\frac{2\Gamma(n+3/2)}{(n+1)!\sqrt{\pi}}U_n(x),
\end{equation}
where $U_n(x)$ is a Chebyshev polynomial of the second kind, as well as
\begin{equation}
    P_n^{(1/2,1/2)}(x)=\frac{\Gamma(n+3/2)}{(n+1)!\Gamma(3/2)}C_{n}^{(1)}(x),
\end{equation}
where $C_n^{(\alpha)}$ represents an ultraspherical Gegenbauer polynomial. In the same way, the so-called generalized characters or order $\lambda$ of the irreducible representation of rank $j$:
\begin{equation}
\chi_{\lambda}^j(\omega)=\frac{\sqrt{2j+1}}{(4j+1)!!}\sqrt{(2j-\lambda)!(2j+\lambda+1)!}2^{2j-\lambda}\left[\sin\left(\frac{\omega}{2}\right)\right]^{\lambda}\,P_{2j-\lambda}^{(\lambda+1/2,\lambda+1/2)}\left[\cos\left(\frac{\omega}{2}\right)\right]
\end{equation}
which can be put in the form
\begin{align}\label{chijlambda}
    \chi_{\lambda}^j(\omega)=&\frac{\sqrt{2j+1}}{(4j+1)!!}\sqrt{\frac{(2j+\lambda+1)!}{(2j-\lambda)!}}2^{2j-\lambda}\left[\sin\left(\frac{\omega}{2}\right)\right]^{\lambda}\left(\lambda+\frac{3}{2}\right)_{2j-\lambda}\nonumber\\
    &\times\,_2F_1\left[\begin{array}{c}
    \lambda-2j,2j+\lambda+2\\
    \lambda+3/2\\
    \end{array};\sin^2\left(\frac{\omega}{4}\right)\right]
\end{align}
or also
\begin{align}
    \chi_{\lambda}^j(\omega)=&\frac{\sqrt{2j+1}}{(4j+1)!!}\sqrt{\frac{(2j+\lambda+1)!}{(2j-\lambda)!}}2^{2j-\lambda}\left[\sin\left(\frac{\omega}{2}\right)\right]^{\lambda}\left(\lambda+\frac{3}{2}\right)_{2j-\lambda}\nonumber\\
    &\times\,f_{2j-\lambda}\left(1,2\lambda+2,2\lambda+2;\lambda+1;-;\sin^2\left(\frac{\omega}{4}\right)\right)
\end{align}
which, for $\lambda=0$, Eq. (\ref{chijlambda}) reduces to Eq. (\ref{chijss}).

\subsection{Rotation matrix and Wigner $d$ functions}

The Wigner $d$ function reads \cite{VARSHALOVICH1988}:
\begin{equation}
    d_{mk}^j(\theta)=\xi_{mk}\left[\frac{s!(s+\mu\nu)!}{(s+\mu)!(s+\nu)!}\right]^{1/2}\left[\sin\frac{\theta}{2}\right]^{\mu}\left[\cos\frac{\theta}{2}\right]^{\nu}\,P_s^{(\mu,\nu)}(\cos\theta)
\end{equation}
with $\mu=|m-k|$, $\nu=|m+k|$, $s=j-(\mu+\nu)/2$, and $\xi_{mk}=1$ if $k\geq m$ and $(-1)^{k-m}$ if $k<m$.

\subsection{Other relations involving Jacobi polynomials}

In 1999, Khan found \cite{KHAN1999}:
\begin{equation}
    \frac{1}{\Gamma(\alpha+1/2)}\int_0^{\infty}t^{\alpha-\frac{1}{2}}\,e^{-t}\,f_n(1,2\alpha+1;-;-;xt)dt=\frac{n!}{(1+\alpha)_n}P_n^{(\alpha,\alpha)}(1-2x),
\end{equation}
as well as
\begin{eqnarray}
    \frac{i}{2\sin\left[(\alpha+\frac{1}{2})\pi\right]\Gamma(\frac{1}{2}-\alpha)}& &\int_{\infty}^{(0+)}(-t)^{-\alpha-\frac{1}{2}}\,e^{-t}\,P_n^{(\alpha,\alpha)}\left(\frac{2x+t}{t}\right)dt\nonumber\\
    &=&\frac{(1+\alpha)_n}{n!}\,f_n(1,2\alpha+1;-;-;x).
\end{eqnarray}

Note that such investigations about the Sister Celine polynomials led to the introduction of generalized Rice polynomials \cite{KHANDEKAR1964}, rediscovered by Khan in 1989. 

\section{Clebsch-Gordan coefficients and Wigner $3j$ symbols}

In quantum mechanics, Clebsch-Gordan coefficients describe how individual angular-momentum states may be coupled to yield the total angular-momentum state of a system. In the literature, Clebsch-Gordan coefficients \cite{VILENKIN1968} are sometimes also known as Wigner coefficients or vector coupling coefficients. They are closely related to Wigner's $3j$ symbol \cite{BIEDENHARN1981} by 
\begin{equation}\label{cle}
C_{a\alpha,~b\beta}^{c\gamma}=(-1)^{a-b+\gamma}\sqrt{2c+1}\threej{a}{b}{c}{\alpha}{\beta}{-\gamma},
\end{equation}
where $C_{a\alpha,~b\beta}^{c\gamma}$ is the Clebsch-Gordan coefficient and 
\begin{equation}\label{3jdb}
\threej{a}{b}{c}{\alpha}{\beta}{\gamma}
\end{equation} 
the $3j$ symbol for angular momenta $a$, $b$ and $c$ with respective projections $\alpha$, $\beta$ and $\gamma$. We have in particular \cite{JEUGT2020}:
\begin{align}\label{3j}
\threej{j_1}{j_2}{j_3}{m_1}{m_2}{m_3}=&(-1)^{j_1-j_2+m_3}\frac{(2j_1)!(j_1+j_2+m_3)!}{(j_1-j_2+j_3)!(j_1+j_2-j_3)!}\nonumber\\
&\times\frac{\Delta(j_1,j_2,j_3)}{\delta(j_1,m_1,j_2,m_2,j_3,m_3)}\nonumber\\
&\times\,_3F_2\left[
    \begin{array}{c}
    m_1-j_1, j_3-j_1-j_2,-j_1-j_2-j_3-1\\
    -2j_1,-j_1-j_2-m_3\\
    \end{array};1
    \right]
\end{align}
with
\begin{equation}
    \Delta(j_1,j_2,j_3)=\sqrt{\frac{(-j_1+j_2+j)!(j_1-j_2+j)!(j_1+j_2-j)!}{(j_1+j_2+j+1)!}}
\end{equation}
and
\begin{equation}
    \delta(j_1,m_1,j_2,m_2,j_3,m_3)=\sqrt{(j_1-m_1)!(j_1+m_1)!(j_2-m_2)!(j_2+m_2)!(j_3-m_3)!(j_3+m_3)!}.
\end{equation}

The Hahn polynomials are a family of orthogonal polynomials in the Askey scheme \cite{ANDREWS1985} of hypergeometric orthogonal polynomials, introduced by Chebyshev in 1875 \cite{CHEBYSHEV1907} and rediscovered by Wolfgang Hahn \cite{HAHN1949,KARLIN1961}. They read \cite{KOEKOEK2010}:
\begin{equation}
Q_n(x;\alpha,\beta,N)=\,_3F_2\left[
    \begin{array}{c}
    -x, -n,n+\alpha+\beta+1\\
    -N, 1+\alpha
    \end{array};1
    \right]
\end{equation}
and the Clebsch-Gordan coefficients can be put in the form \cite{KOORNWINDER1981}:
\begin{align}
    C_{\frac{N}{2}\left(\frac{N}{2}-x\right),~\left(\frac{N+\alpha+\beta}{2}\right)\left(\frac{\alpha-\beta-N}{2}+x\right)}^{\left(n+\frac{\alpha+\beta}{2}\right)\left(\frac{\alpha-\beta}{2}\right)}=&\frac{(-1)^xN!}{\alpha!}\nonumber\\
    &\times\left[\frac{(2n+\alpha+\beta+1)(N-x+\beta)!(x+\alpha)!(n+\alpha)!(n+\alpha+\beta)!}{x!(N-x)!(n+\beta)!n!(N-n)!(N+n+\alpha+\beta+1)!}\right]^{1/2}\nonumber\\
    &\times\,Q_n(x;\alpha,\beta,N),
\end{align}
or with Sister Celine's polynomials
\begin{align}
    C_{\frac{N}{2}\left(\frac{N}{2}-x\right),~\left(\frac{N+\alpha+\beta}{2}\right)\left(\frac{\alpha-\beta-N}{2}+x\right)}^{\left(n+\frac{\alpha+\beta}{2}\right)\left(\frac{\alpha-\beta}{2}\right)}=&\frac{(-1)^xN!}{\alpha!}\nonumber\\
    &\times\left[\frac{(2n+\alpha+\beta+1)(N-x+\beta)!(x+\alpha)!(n+\alpha)!(n+\alpha+\beta)!}{x!(N-x)!(n+\beta)!n!(N-n)!(N+n+\alpha+\beta+1)!}\right]^{1/2}\nonumber\\
    &\times\,f_n(1,1+\alpha+\beta,1+2\alpha;\alpha+\frac{1}{2};-x;-N;1).
\end{align}
One has also \cite{KOORNWINDER1981}:
\begin{align}
    C_{\frac{N}{2}\left(\frac{N}{2}-x\right),~\left(\frac{N+\alpha+\beta}{2}\right)\left(\frac{\alpha-\beta-N}{2}+x\right)}^{\left(N-n+\frac{\alpha+\beta}{2}\right)\left(\frac{\beta-\alpha}{2}\right)}=&{(\alpha+N)!N!}\nonumber\\
    &\times\left[\frac{(2N-2n+\alpha+\beta+1)(n+\beta)!}{x!(N-x)!(\alpha+N-x)!(\beta+x)!}\right]^{1/2}\nonumber\\
    &\times\left[\frac{(n+\alpha+\beta)!}{n!(N-n)!(N-n+\beta)!(2N-n+\alpha+\beta+1)}\right]^{1/2}\nonumber\\
    &\times\,Q_n(x;-N-\alpha-1,-N-\beta-1,N).
\end{align}
which reads, in terms of Sister Celine's polynomials
\begin{align}
    C_{\frac{N}{2}\left(\frac{N}{2}-x\right),~\left(\frac{N+\alpha+\beta}{2}\right)\left(\frac{\alpha-\beta-N}{2}+x\right)}^{\left(N-n+\frac{\alpha+\beta}{2}\right)\left(\frac{\beta-\alpha}{2}\right)}=&{(\alpha+N)!N!}\nonumber\\
    &\times\left[\frac{(2N-2n+\alpha+\beta+1)(n+\beta)!}{x!(N-x)!(\alpha+N-x)!(\beta+x)!}\right]^{1/2}\nonumber\\
    &\times\left[\frac{(n+\alpha+\beta)!}{n!(N-n)!(N-n+\beta)!(2N-n+\alpha+\beta+1)}\right]^{1/2}\nonumber\\
    &\times\,f_n(1,-2N-\alpha-\beta-1,-2N-2\alpha-1,-N-\alpha-\frac{1}{2};-x;-N;1).
\end{align}
Considering the $3j$ coefficient, using the Weber-Erdelyi identity\cite{WEBER1952}:
\begin{align}
_3F_2\left[
    \begin{array}{c}
    -n, \alpha, \beta\\
    \gamma, \delta
    \end{array};1
    \right]=&\frac{\tilde{\Gamma}(\gamma,\gamma+n-\alpha)}{\tilde{\Gamma}(\gamma+n,\gamma-\alpha)}~_3F_2\left[
    \begin{array}{c}
    -n, \alpha, \delta-\beta\\
    1+\alpha-\gamma-n, \delta
    \end{array};1
    \right],
\end{align}
where we introduced the notation $\tilde{\Gamma}(x,y,z,\cdots)=\Gamma(x)\,\Gamma(y)\,\Gamma(z)\cdots$ ($\Gamma$ being the usual Gamma function), we have also (see the notations of Eq. (\ref{3j})) \cite{RAJESWARI1989}:
\begin{align}
    f_n(1,1+\alpha+\beta,1+2\alpha;\alpha+\frac{1}{2};-x;-N;1)=&(-1)^{2j_2+m_1+n+x}\frac{(j_3-j_2-m_1)!}{(2j_2)!}\nonumber\\
    &\times\left[\frac{(j_2-n)!n!(2j_3+n+1)!}{(2j_1-2j_2+n)!(j_3-j_2+m_1+n)!}\right]^{1/2}\nonumber\\
    &\times\left[\frac{x!(2j_2-x)!(j_3-j_2-m_1+n)!}{(j_3-j_2+m_1+x)!(j_3+j_2-m_1-x)!}\right]^{1/2}\nonumber\\
    &\times\threej{j_3-j_2+n}{j_2}{j_3}{m_1}{x-j_2}{j_2-m_1-x},
\end{align}
with
\begin{equation}
\left\{
\begin{array}{l}
n=j_1+j_2-j_3,\\
x=j_2+m_2,\\
N=2j_2+1,\\
\alpha=j_3-j_2+m_1,\\
\beta=-j_2+j_3-m_1.
\end{array}
\right.
\end{equation}
Alternately, it is possible to write \cite{RAJESWARI1989}:
\begin{align}
    f_n(1,1+\alpha+\beta,1+2\alpha;\alpha+\frac{1}{2};-x;-N;1)=&\frac{(-1)^{j_1-j_2-m_3}}{(2j_1)!(j_1+j_2+m_3)!}\nonumber\\
    &\times\left[\frac{(2j_1-n)!n!(2j_1+2j_2-n+1)!}{(2j_2-n)!(j_1+j_2-m_3-n)!}\right]^{1/2}\nonumber\\
    &\times\left[x!(2j_1-x)!(j_1+j_2+m_3-x)!\right]^{1/2}\nonumber\\
    &\times\left[(-j_1+j_2-m_3+x)!(j_1+j_2+m_3-n)!\right]^{1/2}\nonumber\\
    &\times\threej{j_1}{j_2}{j_1+j_2-n}{j_1-x}{-j_1+m_3+x}{m_3},
\end{align}
with
\begin{equation}
\left\{
\begin{array}{l}
n=j_1+j_2-j_3,\\
x=j_1-m_1,\\
N=2j_1+1,\\
\alpha=-j_1-j_2-m_3-1,\\
\beta=-j_1-j_2+m_3-1.
\end{array}
\right.
\end{equation}
Smorodinskii and Suslov obtained the same result \cite{SMORODINSKII1982} using the Weber-Erdelyi transformation \cite{WEBER1952}:
\begin{align}
_3F_2\left[
    \begin{array}{c}
    -n, \alpha, \beta\\
    \gamma, \delta
    \end{array};1
    \right]=&\frac{\tilde{\Gamma}(\gamma,\delta,\delta+n-\alpha,\gamma+n-\alpha)}{\tilde{\Gamma}(\gamma+n,\delta+n,\delta-\alpha,\gamma-\alpha)}\nonumber\\
    &\times~_3F_2\left[
    \begin{array}{c}
    -n, \alpha,\gamma-\delta-n\\
    1+\alpha-\delta-n, 1+\alpha-\gamma-n
    \end{array};1
    \right].
\end{align}
Finally, the following asymptotic connection between Jacobi and Hahn polynomials is worth mentioning \cite{ERDELYI1953,WILSON1970}: 
\begin{equation}
    \lim_{N\rightarrow\infty} Q_n(Nx;\alpha,\beta,N)=\frac{n!}{(\alpha+1)_n}\,P_n^{(\alpha,\beta)}(1-2x).
\end{equation}

\section{Conclusion}

We expressed (generalized) characters of irreducible representations of the rotation group as well as Wigner ``$d$'' matrices in terms of Sister Celine polynomials. This was made possible thanks to expressions of the formers involving Jacobi polynomials. In the second section we recast Clebsch-Gordan coefficients and Wigner $3j$ symbols as Sister Celine's polynomials, by harnessing their expressions in terms of Hahn polynomials. Using known relations, integral or discrete, as well as algorithms for Sister Celine's polynomials, the expressions obtained in the present work should yield new identities, or provide shorter derivations of known formulas related to the quantum theory of angular momentum.

\end{document}